\documentclass[%showpacs,
showkeys,12pt,
preprint,preprintnumbers,nofootinbib,
groupedaddress,superscriptaddress,amsmath,amssymb]{revtex4}
\usepackage{graphicx}% Include figure files
\usepackage{dcolumn}% Align table columns on decimal point
\usepackage{bm}% bold math
\usepackage{amssymb}
\usepackage{amsmath}
\usepackage{epsfig}    
\usepackage{color}
\usepackage{slashed}
\usepackage{hhline}
\usepackage{hyperref}
\hypersetup{colorlinks,bookmarksopen,bookmarksnumbered,citecolor=blue,
linkcolor=black,pdfstartview=FitH,urlcolor=blue}
\def\be{\begin{equation}}
\def\ee{\end{equation}}
\newcommand{\bea}{\begin{eqnarray}}
\newcommand{\eea}{\end{eqnarray}}
\newcommand{\nn}{\nonumber}

\numberwithin{equation}{section}

\begin{document}

%%%%%%%%%
\title{Nonthermal dark matter models and signals}
\preprint{LPT-Orsay-15-80}
\author{Hiroshi Okada}
%\email{hokada@kias.re.kr}
\email{macokada3hiroshi@gmail.com}
\affiliation{School of Physics, KIAS, Seoul 130-722, Korea}
\affiliation{Physics Division, National Center for Theoretical Sciences, Hsinchu, Taiwan 300}

\author{Yuta Orikasa}
\email{orikasa@kias.re.kr}
\affiliation{School of Physics, KIAS, Seoul 130-722, Korea}
\affiliation{Department of Physics and Astronomy, Seoul National
University, Seoul 151-742, Korea}

\author{Takashi Toma}
\email{takashi.toma@th.u-psud.fr}
\affiliation{
Laboratoire de Physique Th\'eorique, CNRS,\\ 
Univ. Paris-Sud, Universit\'e Paris-Saclay, 91405 Orsay, France}

\date{\today}

\begin{abstract}
Many experiments exploring weakly interacting massive particles (WIMPs)
 such as direct, indirect and collider searches have been carried out
 until now. However, a clear signal of a WIMP has not been found yet and it
 makes us to suspect that WIMPs are questionable as a dark matter candidate. 
Taking into account this situation, we propose two models in which dark
 matter relic density is produced by decay of a metastable particle. 
In the first model, the metastable particle is a feebly interacting
 massive particle, which is the so-called FIMP produced by freeze-in
 mechanism in the early universe. 
In the second model, the decaying particle is thermally produced the same
 as the usual WIMP. However decay of the particle into dark matter is
 led by a higher dimensional operator. 
As a phenomenologically interesting feature of nonthermal dark matter discussed
 in this paper, a strong sharp gamma-ray emission as an indirect detection
 signal occurs due to internal bremsstrahlung, although
 some parameter space has already been ruled out by this process. 
 Moreover combining other experimental and theoretical constraints
 such as dark matter relic density, big bang nucleosynthesis, collider,
 gamma-rays and perturbativity of 
 couplings, we discuss the two nonthermal DM models. 

\end{abstract}
\maketitle
\newpage

\section{Introduction}
Exploring the nature of dark matter (DM) is one of the most important issues to
provide an appropriate prescription to improve the standard model (SM). 
The most promising DM candidate is weakly interacting massive particles
(WIMPs) whose mass is predicted to be the order of $10~\mathrm{GeV}$ to
$10~\mathrm{TeV}$, and many experiments are focusing on WIMP searches. 
However in spite of great effort of experiments for WIMP search such as direct, indirect
and collider searches, no positive evidence for WIMPs is found up to the present. 
Although the gamma-ray excess from the galactic center has been
claimed and could be explained by WIMP with its typical
annihilation cross section $\sigma{v}_\mathrm{rel}\sim10^{-26}~\mathrm{cm^3/s}$~\cite{Goodenough:2009gk, Vitale:2009hr, Hooper:2011ti, Abazajian:2014fta, Carlson:2014nra}, 
it is strongly constrained by nondetection of such a gamma-ray excess from the other galaxies. 
In particular, the constraint on the WIMP annihilation cross section
from dwarf spheroidal galaxies is the strongest for specific
channels~\cite{Ackermann:2015zua}. 
For direct detection experiments, the elastic cross section with a nucleon is
strongly constrained, and more and more parameter space of the WIMP
is excluded~\cite{Akerib:2013tjd, Akerib:2015rjg}, 
while this strong bound may be evaded by considering the WIMP interacting with
quarks via a pseudoscalar, leptophilic DM and resonance region in Higgs
portal models. 
Even for collider searches, any collider signal for the WIMP has not been
found yet at the LHC~\cite{Aad:2012fw, ATLAS:2012ky}. 
This may imply that DM in the universe is not composed of the traditional
WIMP candidate, and motivate us to consider non-WIMP DM scenarios. 
There are a lot of DM candidates other than the WIMP, for example
axion~\cite{Visinelli:2009zm, Duffy:2009ig, Baer:2014eja}, asymmetric 
DM~\cite{Kaplan:2009ag, Frandsen:2010yj, Cohen:2010kn}, 
sterile neutrino~\cite{Dodelson:1993je, Shi:1998km}, strongly interacting massive
particle~\cite{Hochberg:2014dra, Hochberg:2014kqa, Bernal:2015bla,
Choi:2015bya}. 

In this paper, we construct two kinds of nonthermal DM
models.\footnote{Some related nonthermal DM production mechanisms have
been discussed in Refs.~\cite{Hooper:2011aj, Mambrini:2013iaa,
Moroi:2013sla, Kelso:2013paa, Chu:2013jja, Kelso:2013nwa, Dev:2013yza,
Blennow:2013jba, Baer:2014eja, Molinaro:2014lfa, Kane:2015qea, Aoki:2015nza}.} 
In both models, the DM particle is produced by decay of a metastable
particle after freeze-out of DM, but the production of the decaying
particle is different. 
Such nonthermally produced DM particles have a phenomenologically interesting feature, 
which is a strong signal for indirect detection. 
For traditional thermally produced DM, the interaction strength of WIMPs
is fixed by the annihilation cross section in 
order to accommodate the correct relic density observed by PLANCK~\cite{Ade:2013zuv}. 
Thus in this case, the signal strength for indirect DM detection is also
determined. 
On the other hand, for nonthermally produced DM like our case, the
strength of the interactions is not fixed and can be larger than the
interaction of WIMPs since the DM relic density is mainly 
generated by the metastable particle decay. 

In the first model, a new decaying particle has only dimension $5$ operators and
the interactions are highly suppressed. Namely this particle can be a
feebly interacting massive particle
(FIMP)~\cite{Hall:2009bx},\footnote{The same mechanism has been
discussed in a concrete model previously~\cite{Asaka:2005cn, Asaka:2006fs}.} 
and is produced in the early universe by so-called freeze-in
scenario. 
The DM particle is nonthermally produced by the decay of FIMP. 
In the second model, both the DM particle and the decaying particle can
be thermally produced at the beginning. 
Then the decaying particle can be metastable since the interactions of
the particle are highly suppressed by dimension $5$ operators. 
The heavier particle decays into the DM particle through the dimension
$5$ operators after DM freeze-out. 
In this way, the DM relic density can be reproduced non-thermally. 
In addition, neutrino masses are generated at one-loop level in the second model. 
We discuss which parameter space in the two models is allowed by some
experimental and theoretical constraints and is favored to see the
nonthermal DM signal. 

This paper is organized as follows. 
In Sec.~\ref{sec:1} and Sec.~\ref{sec:2}, we discuss the first model
(Model I) and the second model (Model II) respectively, in which we
formulate the relevant Lagrangian, the coupled Boltzmann equation for the DM relic
density, neutrino masses, and analyze the DM signature. 
Summary and conclusion are given in Sec.~\ref{sec:3}.

%\newpage

%%%%%%%%%%%%%%%%%%%%%%%%%%%%%%%%%%%%%
\section{The Model I}
\label{sec:1}
\subsection{Model setup}

We consider a model with a discrete symmetry $\mathbb{Z}_4\times\mathbb{Z}_2$. 
The new particle contents and their charge assignments are shown in Table~\ref{tab:1}
where all the SM particles are neutral under the
$\mathbb{Z}_4\times\mathbb{Z}_2$ symmetry. 
These discrete $\mathbb{Z}_N$ symmetries could be understood as a
remnant symmetry of
a $U(1)$ gauge symmetry which comes from string
theory~\cite{BerasaluceGonzalez:2011wy}. 
As in Table ~\ref{tab:1}, we add two gauge singlet right-handed fermions $X$
and $N$, a charged singlet scalar $S^+$ and a neutral singlet scalar
$S^0$ to the SM. 
%%% 
The kinetic terms of the new particles and the Majorana mass terms of
the new fermions are given by 
\begin{equation}
\mathcal{L}_K=
\frac{1}{2}\overline{X^c}\left(i\partial\hspace{-0.22cm}/-m_X\right)X
+\frac{1}{2}\overline{N^c}\left(i\partial\hspace{-0.22cm}/-m_N\right)N
+\left|\partial_\mu S^0\right|^2
+\left|D_\mu S^+\right|^2,
\end{equation}
where the covariant derivative for $S^+$ is defined by
$D_\mu\equiv\partial_\mu+ig_YB_\mu$ with the $U(1)_Y$ gauge coupling
$g_Y$ and the $U(1)_Y$ field $B_\mu$. 
Under the charge assignment, the relevant Lagrangian for Yukawa sector
up to dimension 5 operators is given by
\begin{eqnarray}
\mathcal{L}_{Y}&=&
 -y_N S^+\overline{N^c}e_R  -  y_\ell H\overline{L_L}e_R
- \frac{\lambda_\nu}{4\Lambda}(HH\overline{L^c_L}L_L)\nonumber\\
&& 
 - \frac{\lambda_1}{2\Lambda}(\overline{X^c} X)|H|^2 
  - \frac{\lambda_2}{2\Lambda}(\overline{N^c} N)|H|^2 
  - \frac{\lambda_3}{2\Lambda}(\overline{X^c}N)({S^0}^\dag)^2
  - \frac{\lambda_4}{2\Lambda}(\overline{X^c}X)|S^0|^2\nonumber\\
&&   - \frac{\lambda_5}{2\Lambda}(\overline{X^c}X)|S^+|^2
  - \frac{\lambda_6}{2\Lambda}(\overline{N^c}N)|S^0|^2
- \frac{\lambda_7}{2\Lambda}(\overline{N^c}N)|S^+|^2+\mathrm{H.c.},
\label{eq:yukawa}
\end{eqnarray}
where $\Lambda$ is a cutoff scale of the model, $H$ is the Higgs
doublet, $L_L$ and $e_R$ are the $SU(2)_L$ doublet 
and singlet SM lepton fields.\footnote{Notice here that there exists a dimension
5 operator $\overline{X} \sigma^{\mu\nu}X F_{\mu\nu}$ if the fermion $X$
is a Dirac field.} 
In general, the Yukawa coupling $y_N$ is possible for all the leptons,
however we consider the dominant coupling with electron for simplicity.
The charged lepton masses can be induced by the term $ y_\ell
\overline{L_L} H e_R$ as same as the SM, and 
the neutrino masses can be generated by the Weinberg
operator with the $\lambda_\nu$ coupling in
Eq.~(\ref{eq:yukawa})~\cite{Weinberg:1980bf}. 
From the Weinberg operator, the cutoff scale $\Lambda$ is estimated as
$\Lambda\sim10^{14}\lambda_\nu~\mathrm{GeV}$ where the neutrino mass scale is assumed
to be $m_\nu\sim0.1$ eV.

\begin{widetext}
\begin{center} 
\begin{table}[t]
\begin{tabular}{|c||c|c||c|c|}\hline
& $X$& $N$ & $S^0$ & $S^+$ \\\hline 
$(SU(2)_L,U(1)_Y) $ & $(\bm{1},0)$ & $(\bm{1},0)$ & $(\bm{1},0)$  &$(\bm{1},1)$\\\hline 
$(\mathbb{Z}_4,\mathbb{Z}_2)$  & $(-1,-)$   & $(+1,-)$ & $(\pm i,+)$  &
 $(+1,-)$  \\\hline
Spin & 1/2 & 1/2 & 0 & 0 \\\hline
%%%
\end{tabular}
\caption{New particle contents of Model I and their charge assignments under
 $SU(2)_L\times U(1)_Y\times \mathbb{Z}_4\times\mathbb{Z}_2$, where all
 the SM particles are neutral under $\mathbb{Z}_4\times\mathbb{Z}_2$. 
}
\label{tab:1}
\end{table}
\end{center}
\end{widetext}

Only the SM Higgs field denoted as $H$ and the new singlet
scalar $S^0$ should have vacuum
expectation values (VEVs), which are symbolized by $\langle H\rangle=v/\sqrt{2}\approx
174$ GeV and $\langle S^0\rangle=v'/\sqrt{2}$ respectively. 
The $ \mathbb{Z}_4$ symmetry is spontaneously broken by the VEV of
$S^0$, whereas the $\mathbb{Z}_2$ symmetry remains even after the
electroweak symmetry breaking. Hence the $\mathbb{Z}_2$ symmetry assures the
stability of DM, and we can identify the Majorana fermion $X$ or $N$ as
a DM candidate since they are neutral and have the $\mathbb{Z}_2$ odd charge. 
The Majorana fermions $X$ and $N$ mix with each other due to the VEV of $S^0$ via the
coupling $\lambda_3$, and the mixing mass is given by
$m_{XN}=\lambda_3{v'}^2/(2\Lambda)$. However since the mixing occurs
with the dimension 5 operator and the cutoff scale $\Lambda$ is expected to
be very large in order to obtain the appropriate neutrino mass scale with
$\mathcal{O}(1)$ dimensionless couplings $\lambda_\nu$, 
we can naturally expect that the mixing component is very small compared
to the diagonal elements and one can regard that the Majorana fermions
$X$ and $N$ are almost mass eigenstates
themselves. 
The SM Higgs boson $H^0$ and $S^0$ mix after the electroweak symmetry
breaking and the mixing angle is constrained by experiments~\cite{Robens:2015gla}. 
However this is not relevant to our work.

\subsection{Dark matter}
\subsubsection{Freeze-in scenario}
In this model, one of the Majorana fermions $X$ and $N$ can be a DM
candidate depending on the mass hierarchy. 
Since all the interactions of the fermion $X$ are suppressed by the
cutoff scale $\Lambda$ as one can see from Eq.~(\ref{eq:yukawa}), the
fermion $X$ may not be suitable as a 
standard thermally produced DM candidate. 
However, because of the highly suppressed interactions, the fermion $X$
may never reach to thermal equilibrium with the SM particles. 
In this case, the production of
the fermion $X$ occurs by so-called freeze-in mechanism~\cite{Hall:2009bx, Blennow:2013jba}.
Although the nonthermally produced $X$ itself can be a DM candidate, it would be
difficult to search such a DM candidate since $X$ has only
extremely suppressed interactions. 
The most phenomenologically interesting possibility would be a scenario that the
Majorana fermion $N$ is the DM candidate which is reproduced by the decay of the fermion $X$
after the DM freeze-out. 
Because of the nonthermal production mechanism of DM $N$, 
one can expect a larger indirect detection signal of DM since
the interactions of nonthermally produced DM can generally be larger than 
traditional thermally produced DM. 
Thus we will discuss this scenario below.

The following coupled Boltzmann equation for $N$ and $X$ should be solved in
order to compute the DM relic density
%Let us discuss in details. Each of Boltzmann equation for $X_R$ and $N_R$ is given by 
%\begin{eqnarray}
%\frac{dn_X}{dt}+3Hn_X&=&\frac{g_Xm_X^2m_N\Gamma_X}{2\pi^2x}K_1\left(\frac{m_X}{m_N}x\right)-\Gamma_Xn_X\nn\\
%\frac{dn_N}{dt}+3Hn_N&=&-\langle\sigma_N{v_{\rm rel}}\rangle\left(n_N^2-{n_N^{\mathrm{eq}}}^2\right)
%+N_\mathrm{dec}\Gamma_Xn_X,
%\end{eqnarray}
%or 
\begin{eqnarray}
\frac{dY_X}{dx}\hspace{-0.1cm}&=&\hspace{-0.1cm}
\frac{1}{sxH}\left(\frac{g_Xm_X^2m_N\Gamma_X}{2\pi^2x}\right)K_1\left(\frac{m_X}{m_N}x\right)
-\frac{\Gamma_XY_X}{xH}, \nn\\
\frac{dY_N}{dx}\hspace{-0.1cm}&=&\hspace{-0.1cm}
-\frac{s\langle\sigma_\mathrm{eff}{v_{\rm rel}}\rangle}{xH}\left(Y_N^2-{Y_N^\mathrm{eq}}^2\right)
+\frac{\Gamma_XY_X}{xH},
\label{eq:bltm-Y}
\end{eqnarray}
where $g_X=2$ is the degree of freedom of the Majorana fermion $X$, $x=m_N/T$ is a
dimensionless parameter with the temperature of the universe, 
$Y_N$ and $Y_X$ are 
defined by $Y_N\equiv n_N/s$ and $Y_X\equiv n_X/s$ with the number
densities $n_N$, $n_X$ and the entropy density $s$, 
$Y_N^{\mathrm{eq}}$ represents the number density of $N$ in thermal equilibrium, 
$H$ is the Hubble parameter, and $K_1$ is the modified Bessel function of
the second kind with the order 1. 
The first term including the modified Bessel function in
Eq.~(\ref{eq:bltm-Y}) implies the $X$ production due to the inverse decay
process $NS^0\to X$ via the dimension $5$ operator with $\lambda_3$
where $m_X> m_N+m_{S^0}$ is assumed. 
One may think that the scattering processes induced by the other dimension
$5$ operators in Eq.~(\ref{eq:yukawa}) should {also} be taken into account
and be added to the coupled Boltzmann equation. 
In fact if the reheating temperature of the universe is high enough compared
to the following criterion Eq.~(\ref{eq:criterion}), 
the time evolution of the number density of $X$ is dominantly
determined by the scattering processes. 
While if the reheating temperature is not so high, the time evolution is
almost determined by the inverse decay process we included. 
Thus the assumption that the inverse decay process is dominant for
freeze-in mechanism gives a constraint on reheating temperature. 
The constraint on the reheating temperature is roughly estimated as~\cite{Hall:2009bx} 
\begin{equation}
T_R\lesssim \frac{3\pi^2 v'^2}{m_X}.
\label{eq:criterion}
\end{equation}
For example, when $m_X=10~\mathrm{TeV}$, $v'=3~\mathrm{TeV}$, the upper
bound of the reheating temperature is derived as
$T_R\lesssim27~\mathrm{TeV}$. 
More general analysis for treatment of nonrenormalizable operators has
been discussed in Ref.~\cite{Elahi:2014fsa}. 

The fermion $X$ can decay as $X\to NS^0$ via the coupling $\lambda_3$, and the decay
width $\Gamma_X$ is computed as 
\begin{equation}
\Gamma_X=\frac{\lambda_3^2m_X}{16\pi}
\left(\frac{v'}{\Lambda}\right)^2
\sqrt{1-\left(\frac{m_N}{m_X}+\frac{m_{S^0}}{m_X}\right)^2}
\sqrt{1-\left(\frac{m_N}{m_X}-\frac{m_{S^0}}{m_X}\right)^2}
\left[\left(1-\frac{m_N}{m_X}\right)^2-\frac{m_{S^0}^2}{m_X^2}\right].
\label{eq:dw-X}
\end{equation}
Thus one can obtain the rough estimation for $m_X\gg m_N,m_{S^0}$ as
\begin{equation}
\Gamma_X\sim
2\times10^{-20}
\left(\frac{\lambda_3}{1}\right)^2
\left(\frac{m_X}{10~\mathrm{TeV}}\right)
\left(\frac{v'}{1~\mathrm{TeV}}\right)^2
\left(\frac{10^{14}~\mathrm{GeV}}{\Lambda}\right)^2~\mathrm{GeV}.
\end{equation}

The DM annihilation cross section $\sigma_{NN}{v_{\rm rel}}$ can be
expanded by the DM relative velocity $v_\mathrm{rel}$ as usual way. 
In this model, the DM annihilation channel is only $NN\to
e_R\overline{e_R}$ via the Yukawa coupling $y_N$, and the concrete
expression of the expansion is given by
\begin{equation}
\sigma_{NN}{v_{\rm rel}}=
\frac{y_N^4}{48\pi m_N^2}\frac{1+\mu_N^2}{\left(1+\mu_N\right)^4}v_{\rm rel}^2,
\label{eq:p-wave}
\end{equation}
with $\mu_N=m_{S^+}^2/m_N^2$. 
The first term of the expansion which corresponds to $s$-wave does not exist because
of chiral suppression. 
The thermally averaged cross section $\langle\sigma_{NN}{v_{\rm rel}}\rangle$
is given by replacing $v_\mathrm{rel}^2\to6/x$ in Eq.~(\ref{eq:p-wave}).
In addition to the DM annihilation, the coannihilation processes with
$S^\pm$ should be taken into account since we will consider the degenerate
mass $m_N\approx m_{S^+}$ in order to obtain an interesting DM signal in indirect detection. 
The coannihilation cross sections and self-annihilation cross sections of
$S^\pm$ for each channel are computed as
\begin{eqnarray}
\sigma{v}_\mathrm{rel}(S^+S^-\to\gamma\gamma)&=&
\frac{e^4}{8\pi m_N^2}\left(1-\frac{7}{12}v_\mathrm{rel}^2\right),
\label{eq:sann1}\\
\sigma{v}_\mathrm{rel}(S^+S^-\to \gamma
 Z)&=&\frac{e^4\tan^2\theta_W}{4\pi
 m_N^2}\left(1-\frac{7}{12}v_\mathrm{rel}^2\right),
\label{eq:sann2}\\
\sigma{v}_\mathrm{rel}(S^+S^-\to ZZ)&=&\frac{e^4\tan^4\theta_W}{8\pi
 m_N^2}\left(1-\frac{7}{12}v_\mathrm{rel}^2\right),
\label{eq:sann3}\\
\sigma{v}_\mathrm{rel}(S^+S^-\to W^+W^-)&=&\frac{e^4}{1536\pi
 m_N^2}\frac{m_Z^4}{m_W^4}v_\mathrm{rel}^2,
\label{eq:sann4}\\
\sum_{f}\sigma{v}(S^+S^-\to
 f\overline{f})&=&\left(\frac{1}{768}\frac{5e^4}{\cos^4\theta_W}-\frac{1}{96}\frac{y_N^2e^2}{\cos^2\theta_W}+\frac{y_N^4}{192}\right)\frac{v_\mathrm{rel}^2}{\pi
 m_N^2}
\label{eq:ssff}
,\\
\sigma{v}_\mathrm{rel}(S^\pm S^\pm\to e^\pm e^\pm)&=&
\frac{y_N^4}{16\pi m_N^2}\left(1-\frac{v_\mathrm{rel}^2}{3}\right),
\label{eq:ss}\\
\sigma{v}_\mathrm{rel}(S^\pm N\to e^\pm\gamma)&=&
\frac{y_N^2e^2}{64\pi m_N^2}\left(1-\frac{v_\mathrm{rel}^2}{4}\right),
\label{eq:coann1}\\
\sigma{v}_\mathrm{rel}(S^\pm N\to e^\pm
 Z)&=&\frac{y_N^2e^2\tan^2\theta_W}{64\pi
 m_N^2}\left(1-\frac{v_\mathrm{rel}^2}{4}\right),
\label{eq:coann2}
%\\
%\sigma{v}_\mathrm{rel}(S^\pm N\to \nu W^\pm)&\propto&\frac{m_e^2}{m_W^2}\to0,
\end{eqnarray}
where $f$ in Eq.~(\ref{eq:ssff}) represents the SM fermions, the SM Yukawa
couplings and the $|S^+|^2|H|^2$ coupling are 
neglected for simplicity, and the mass relations $m_W,m_Z\ll m_N$
and $\mu_N=1$ are assumed. 
The co-annihilation cross section for the process $S^\pm N\to\nu W^\pm$
is proportional to $m_e^2/m_W^2$. In addition, the cross section for $S^\pm N\to
he^\pm$ is written by the electron Yukawa coupling where $h$ is the
SM-like Higgs boson with $m_h=125$ GeV. Thus these contributions are 
negligible.  
We have computed the above analytical formulas with
FEYNCALC~\cite{Mertig:1990an}, 
and have numerically checked with CALCHEP~\cite{Pukhov:1999gg, Pukhov:2004ca}. 
The general formula of the effective cross section including
coannihilation processes is given by~\cite{Griest:1990kh}
\begin{equation}
\sigma_\mathrm{eff}{v}_\mathrm{rel}=
\sum_{i,j}\frac{g_ig_j}{g_\mathrm{eff}^2}\sigma_{ij}v_\mathrm{rel}
\left(1+\Delta_i\right)^{3/2}\left(1+\Delta_j\right)^{3/2}
e^{-x(\Delta_i+\Delta_j)},
\end{equation}
where $i,j$ imply the DM particle ($N$) and the degenerate particles
with DM ($S^\pm$), $\Delta_i\equiv(m_i-m_N)/m_N$, $g_i$ is the degree of freedom
of the particle $i$ and the effective degree of freedom $g_\mathrm{eff}$
is given by
\begin{equation}
g_\mathrm{eff}=\sum_{i}g_i\left(1+\Delta_i\right)^{3/2}e^{-x\Delta_i}.
\end{equation} 
In our case with $\mu_N=1$, the effective cross section including all the
above processes is simply given by 
\begin{equation}
\sigma_\mathrm{eff}{v}_\mathrm{rel}=
\frac{\sigma_{NN}v_\mathrm{rel}}{4}
+\frac{\sigma_{S^\pm S^\mp}v_\mathrm{rel}}{8}
+\frac{\sigma_{S^\pm S^\pm}v_\mathrm{rel}}{8}
+\frac{\sigma_{NS^\pm}v_\mathrm{rel}}{2},
\label{eq:eff_sv}
\end{equation}
where $\sigma_{S^\pm S^\mp}v_\mathrm{rel}$ is defined by the sum of Eq.~(\ref{eq:sann1}),
(\ref{eq:sann2}), (\ref{eq:sann3}), (\ref{eq:sann4}) and
(\ref{eq:ssff}), $\sigma_{S^\pm S^\pm}v_\mathrm{rel}$ is the
contribution of $S^\pm S^\pm\to e^\pm e^\pm$ given by
Eq.~(\ref{eq:ss}), and $\sigma_{NS^\pm}v_\mathrm{rel}$ is given by
the sum of Eq.~(\ref{eq:coann1}) and (\ref{eq:coann2}). 
The thermally averaged effective cross section
$\langle\sigma_\mathrm{eff}v_\mathrm{rel}\rangle$ is needed to solve the
Boltzmann equation Eq.~(\ref{eq:bltm-Y}).

%%%% %%%
\subsubsection{Numerical result}
\begin{figure}[tcb]
\begin{center}
\includegraphics[scale=0.65]{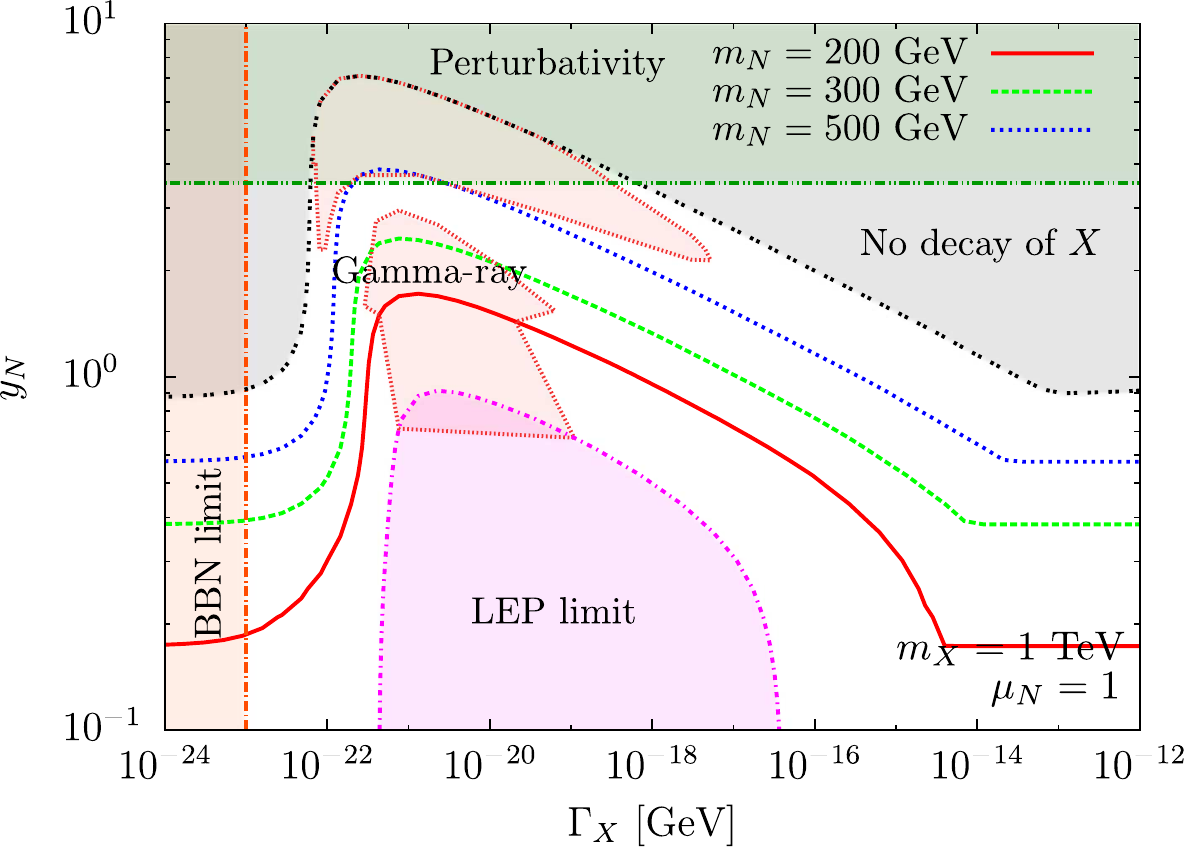}
\includegraphics[scale=0.65]{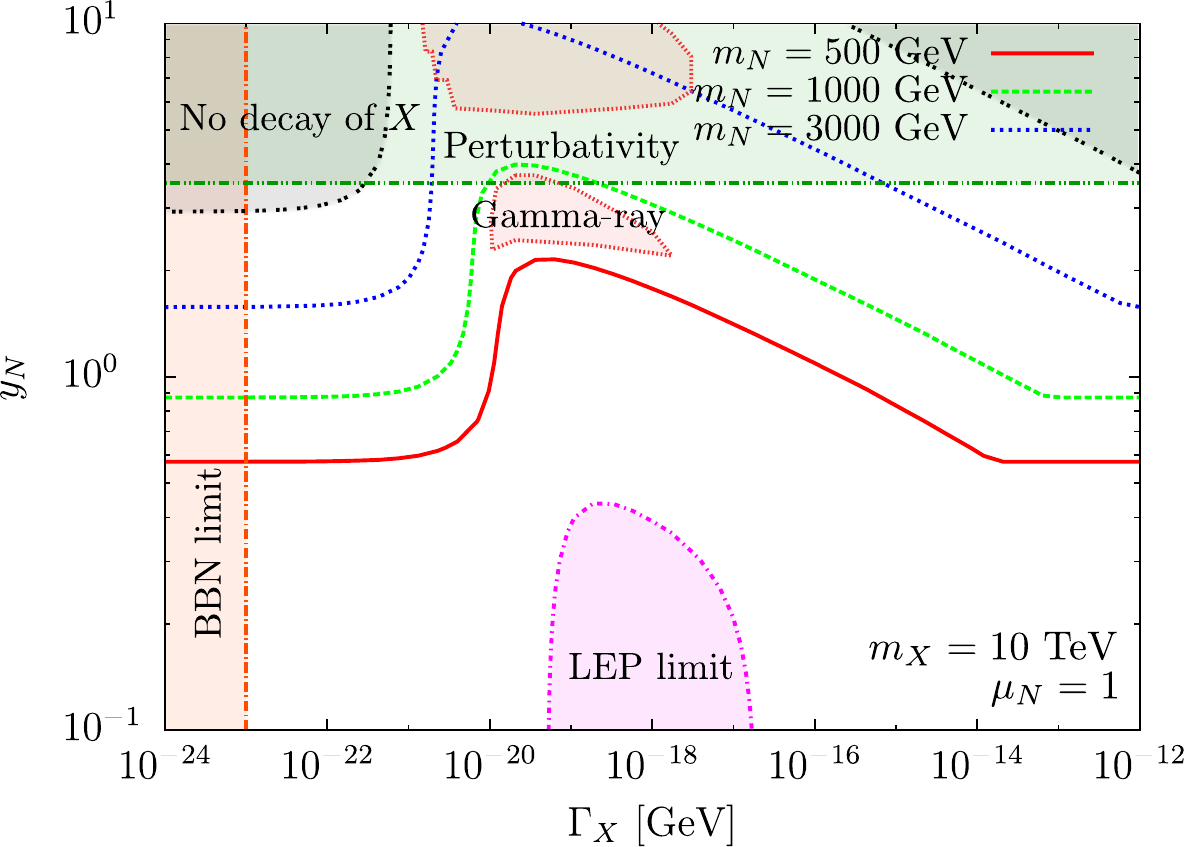}
   \caption{Allowed parameter space in $\Gamma_X$-$y_N$ plane where the
 mass ratio is fixed as $\mu_N=1$ and
 the mass of the Majorana fermion $X$ is taken to be $m_X=1$ TeV in the
 left panel and to be $m_X=10$ TeV in the right panel. 
 The red, green, and blue colored lines imply the contours satisfying the
 observed relic density $\Omega h^2\approx0.12$ for fixed DM masses. 
 The white region is allowed by all the current experimental and
 theoretical bounds.} 
   \label{fig:Ver.I}
\end{center}
\end{figure}

The coupled Boltzmann equation in Eq.~(\ref{eq:bltm-Y}) combined with the $X$ decay
width Eq~(\ref{eq:dw-X}) and the DM cross section Eq.~(\ref{eq:eff_sv}),
is numerically solved. 
Figure~\ref{fig:Ver.I} shows the numerical results for
$\mu_N=1$ in $\Gamma_X$-$y_N$ plane where the decaying Majorana fermion mass $m_X$
is fixed to $m_X=1~\mathrm{TeV}$ in the left panel and $10~\mathrm{TeV}$
in the right panel. 
Each red, green and blue colored line satisfies the observed
relic density $\Omega h^2\approx0.12$ for the fixed DM mass
$m_N=200,300,500$ GeV in the left panel and $m_N=500,1000,3000$ GeV
in the right panel respectively. 

The black colored region represents $m_N>m_X$, thus the decay of $X$ does not occur. 
The green colored upper region is excluded by the conservative
perturbativity of the Yukawa coupling $y_N\geq\sqrt{4\pi}$. 
If the lifetime of $X$ is as long as $\tau_X\sim0.1~\mathrm{s}$
corresponding to $\Gamma_X\sim10^{-23}~\mathrm{GeV}$, the $X$ decay
may affect to the successful big bang
nucleosynthesis (BBN)~\cite{Kawasaki:2004qu, Jedamzik:2006xz}. Therefore the 
conservative limit for the lifetime $\tau_X\lesssim0.1~\mathrm{s}$ is
imposed in our analysis, and the left orange region of
Fig.~\ref{fig:Ver.I} shown with BBN limit is excluded by this constraint. 
%

%%%%%%%%%%%%%%%%%%%

The light-red and violet colored regions in the center of each figure are excluded by
the gamma-ray and LEP experiment respectively~\cite{LEP_bound, Fox:2011fx}. 
For the LEP bound, we take a conservative lower bound for the charged scalar
$m_{S^+}\geq100~\mathrm{GeV}$ which corresponds to
$m_{N}\geq100~\mathrm{GeV}$ since the mass ratio is fixed to be
$\mu_N=1$. 
%This bound is shown as the violet region with LEP limit in
%Fig.~\ref{fig:Ver.I}. 
%
For the gamma-ray constraint, we take into account internal
bremsstrahlung of Majorana DM $NN\to e\overline{e}\gamma$~\cite{Bergstrom:1989jr,
Flores:1989ru, Bringmann:2007nk, Ciafaloni:2011sa, Barger:2011jg,
Garny:2013ama, Kopp:2014tsa, Okada:2014zja}.\footnote{Internal bremsstrahlung has also
been discussed for scalar DM coupling with a vectorlike
fermion~\cite{Toma:2013bka, Giacchino:2013bta, Ibarra:2014qma,
Giacchino:2014moa}. In this case, further strong gamma-ray emission is expected due to
stronger $d$-wave suppression for 2-body annihilation cross section.} 
Indeed in our case, this process is promising channel for indirect detection of
DM, since the DM self-annihilation channel is
chirally suppressed with $v_\mathrm{rel}\sim10^{-3}$ as in Eq.~(\ref{eq:p-wave}). 
For the case of thermal DM, this cross section is fixed to obtain the
observed DM thermal relic density, and cannot be so large. 
However in our nonthermal DM model, it can be large enough to be
detectable in the near future since we can take a larger
Yukawa coupling $y_N$ being consistent with the observed DM relic density. 
The gamma-ray spectrum coming from internal bremsstrahlung $NN\to
e\overline{e}\gamma$ especially becomes very sharp when the mass ratio $\mu_N$ is 
close to $1$ and may give a strong constraint on our model. 
That is why the mass ratio $\mu_N$ is fixed to be $\mu_N=1$ in our analysis
in order to obtain a sharp gamma-ray spectrum of internal bremsstrahlung.
The total cross section for the process is given by
\begin{equation}
\sigma_{e\overline{e}\gamma}v_\mathrm{rel}=\frac{\alpha_\mathrm{em}y_N^4}{64\pi^2 m_N^2}
\left(\frac{7}{2}-\frac{\pi^2}{3}\right),
\end{equation}
with the mass ratio $\mu_N=1$. 
This cross section is constrained by the current gamma-ray experiments
such as Fermi-LAT~\cite{Ackermann:2015zua} and
H.E.S.S.~\cite{Abramowski:2013ax}, and we take the bound 
which has been obtained in Refs.~\cite{Garny:2013ama, Garny:2015wea}. 
The target energy range is 40 GeV to 300 GeV for Fermi-LAT and $500$ GeV
to 25 TeV for H.E.S.S.. 
The bound has been obtained by performing a binned profile likelihood
analysis and assuming the Einasto
profile with the local DM density $\rho_\odot=0.4~\mathrm{GeV/cm^3}$. 
The data of the gamma-ray flux have been taken from search region 3,
Pass7 SOURCE sample for Fermi-LAT as described in Ref.~\cite{Weniger:2012tx}, and from CGH
region for H.E.S.S.~\cite{Abramowski:2013ax} with the expected energy
resolution of Fermi-LAT and H.E.S.S. respectively.
As mentioned in Ref.~\cite{Garny:2013ama}, the 43 months Fermi-LAT data
and the 112h H.E.S.S. data have been analyzed in order to get the
bound. 
Although a similar sharp spectrum of $e^+e^-$ is induced and the model may be
constrained by the $e^+e^-$ measurement of AMS-02~\cite{Aguilar:2013qda}, this is
much weaker than the gamma-ray constraint and does not give a
substantial bound.

Here we notice that deviation from $\mu_N=1$ may weaken the constraint of the
gamma-ray in the central region in Fig.~\ref{fig:Ver.I} because the
energy spectrum of gamma-ray
coming from internal bremsstrahlung becomes broad. 
Simultaneously the strong gamma-ray signature of nonthermal DM may not
be visible. 
However another constraint from the LHC arises through the $S^\pm$
production as follows. 
A pair of the charged scalar $S^\pm$ is produced at the LHC and they
decay into $S^\pm\to e^\pm N$ via the Yukawa coupling $y_N$. 
This decay width becomes large enough to decay inside the detector if
the mass splitting between $S^\pm$ and the DM particle $N$ given by the parameter
$\mu_N$ deviates from $\mu_N=1$. 
As a result, a nontrivial constraint would be obtained, 
but the situation is beyond our scope. 
The lower bound for the DM mass obtained from the LHC can be roughly estimated as
$m_N\gtrsim300~\mathrm{GeV}$ from analogy with the analysis for slepton
search in supersymmetric models at the LHC~\cite{TheATLAScollaboration:2013hha}.

The white region in Fig.~\ref{fig:Ver.I} is allowed by all the  
current experimental and theoretical constraints. 
From the figure, one can see the allowed region of the $X$ decay width
inducing a large Yukawa coupling $y_N$ for the sharp gamma-ray
of internal bremsstrahlung is roughly estimated as 
\begin{eqnarray}
&&
10^{-22}~\mathrm{GeV}\lesssim\Gamma_X\lesssim10^{-16}~\mathrm{GeV}
\mbox{\hspace{0.3cm}for\hspace{0.3cm}}
m_X=1~\mathrm{TeV},\\
&&
10^{-21}~\mathrm{GeV}\lesssim\Gamma_X\lesssim10^{-15}~\mathrm{GeV}
\mbox{\hspace{0.3cm}for\hspace{0.3cm}}
m_X=10~\mathrm{TeV}.
\end{eqnarray}
Thus one can read off the promising parameter region of 
$\epsilon_1\equiv\lambda_3v'/\Lambda$ using Eq.~(\ref{eq:dw-X}) as 
\begin{equation}
2.2\times10^{-12}\lesssim\epsilon_1\lesssim2.2\times10^{-9}
\mbox{\hspace{0.3cm}for\hspace{0.3cm}}
m_X=1,10~\mathrm{TeV}.
\end{equation}

%%%%%%%%%%%%%%%%%%%%%%
\section{Model II}
\label{sec:2}
\subsection{Model setup}
Next we discuss Model II, in which the new particle contents and their
charge assignments are shown in Table~\ref{tab:2}. 
In addition to the particle contents of Model I which have been discussed
in the previous section, we further add one $SU(2)_L$ doublet 
inert boson $\eta$, and the $\mathbb{Z}_8$ symmetry is imposed instead of
the $\mathbb{Z}_4$ symmetry in Model I. 
This $\mathbb{Z}_8$ symmetry is spontaneously broken by the VEV of
$S^0$, but $\mathbb{Z}_2$ symmetry is the exact symmetry even after 
the electroweak symmetry breaking. Hence the $\mathbb{Z}_2$ symmetry assures the
stability of DM like Model I. 
%%% 
We assume that only the SM Higgs denoted as $H$ and the scalar $S^0$
have VEVs symbolized by $\langle H\rangle=v/\sqrt2$, $\langle
S^0\rangle=v'/\sqrt2$ respectively. 

The relevant Lagrangian up to dimension $5$ operators 
under the above charge assignment is given by
\begin{eqnarray}
\mathcal{L}
&\supset&
-\frac{y_X}{2} {S^0}^\dag\overline{X^c}X - \frac{y_S}{2} {S^0}\overline{N^c}N
-y_N S^+\overline{N^c}e_R - y_\ell H\overline{L_L}e_R -  y_\eta
\eta^\dag\overline{L_L}X\nonumber\\
&&
-\frac{\lambda_{H\eta}}{2}(H^\dag\eta)^2
-\frac{1}{2\Lambda}\left(\xi_S{S^{0}}^2+\xi_S'{S^{0\dag}}^2\right)\overline{N^c}X
-\frac{1}{2\Lambda}\left(\kappa_S{S^0}^2+\kappa_S'{S^{0\dag}}^2\right)(\eta
H)S^- +{\rm H.c.}.\nonumber\\
\end{eqnarray}
The VEV of $S^0$ gives the masses of the Majorana
fermions $X$ and $N$ which are symbolized by $m_{N}\equiv y_S v'/\sqrt2$
and $m_{X}\equiv y_\chi 
v'/\sqrt2$. 
The same as Model I, we assume that the Yukawa couplings $y_N$ and
$y_\eta$ only couple with electron for simplicity, and we can naturally
expect that the Majorana
fermions $X$ and $N$ are almost mass eigenstates since the mixing is
generated by the small dimension $5$ operators of the $\xi_S$ and $\xi_S'$
terms. 

\begin{center} 
\begin{table}[tbc]
\begin{tabular}{|c||c|c|c|c||c|c|c|c|}\hline
& $L_L$& $e_R$ & $X$ & $N$ & $H$ & $\eta$ & $S^+$ & $S^0$\\\hline 
$(SU(2)_L,U(1)_Y) $ & $(\bm{2},-1/2)$ & $(\bm{1},-1)$ & $(\bm{1},0)$
 &$(\bm{1},0)$ & $(\bm{2},1/2)$ & $(\bm{2},1/2)$ & ($\bm{1},1$) &
 ($\bm{1},0$)\\\hline  
($\mathbb{Z}_{8},~\mathbb{Z}_2$)  & $(1,~+)$   & $(1,~+)$ & $(1,~-)$  & $(3,~-)$ & $(0,~+)$ & $(0,~-)$
 & $(4,~-)$ & $(2,~+)$\\\hline
Spin & 1/2 & 1/2 & 1/2 & 1/2 & 0 & 0 & 0 & 0\\\hline
\end{tabular}
\caption{Particle contents of Model II and their charge assignments under $SU(2)_L\times
 U(1)_Y\times \mathbb{Z}_8\times\mathbb{Z}_2$. 
}
\label{tab:2}
% \end{tiny}
\end{table}
\end{center}

{\it Higgs sector}:\\
%%%
Although the CP even neutral scalars with nonzero VEVs ($H,S^0)$ mix with each
other the same as Model I, the mixing is not relevant with the following analysis. 
The charged scalars ($\eta^+,S^+$) also mix with each other through the
dimension $5$ operators including $\kappa_S,\kappa_S'$, and this mixing 
plays an important role in nonthermal DM production since this mixing
leads the decay of $X$ into DM $N$. 
The charged scalars $\eta^+$ and $S^+$ are rewritten in
terms of the mass eigenstates $H_1^+$ and $H_2^+$ as 
\begin{align}
\eta^+ &= H_1^+\cos\theta - H_2^+\sin\theta,\nn\\ 
S^+ &= H_1^+\sin\theta + H_2^+\cos\theta,
\label{eq:mass_weak}
\end{align}
where the mixing angle $\theta$ is given by
\begin{equation}
\sin 2\theta=\frac{2\epsilon_2vv'}{m^2_{H_1}-m_{H_2}^2},\quad
\mbox{with}\quad 
\epsilon_2\equiv
\frac{\left(\kappa_{S}+\kappa_{S}'\right)v'}{4\sqrt2\Lambda}. 
\end{equation}

{\it Lepton sector}:\\
%%%
The Weinberg operator $HH\overline{L_L^c}L_L/\Lambda$ is forbidden by the $\mathbb{Z}_8$
symmetry in this model. 
However, the neutrino masses can be derived at the one-loop
level like the Ma model~\cite{Ma:2006km}. 
The neutrino mass formula is given by 
\begin{align}
(m_{\nu})_{\alpha\beta}=\sum_{i} \frac{(y_\eta)_{\alpha i}(y_\eta)_{\beta i} m_{X_i}}{2(4\pi)^2}
\left[
\frac{m^2_R}{m^2_R-m^2_{X_i}}\ln\left(\frac{m^2_R}{m^2_{X_i}}\right)
-
\frac{m^2_I}{m^2_I-m^2_{X_i}}\ln\left(\frac{m^2_I}{m^2_{X_i}}\right)
\right],
\label{eq:neutrino}
\end{align}
where each of $m_R$ and $m_I$ is the mass eigenvalue of the inert
neutral component of the doublet scalar $\eta$; $\eta_R$ and $\eta_I$,
which is defined in Ref.~\cite{Ma:2006km}. 
The mass difference between $\eta_R$ and $\eta_I$ is given by
$m_R^2-m_I^2=\lambda_{H\eta}v^2$, which is essential to 
generate nonzero neutrino masses as one can see from the above mass formula. 
Note that if one requires to reproduce the neutrino oscillation data correctly, 
at least two kinds of the Majorana fermions $X_i$ are needed as denoted
by $i$ in Eq.~(\ref{eq:neutrino}). 
In addition, the constraints from lepton flavor violating processes such
as $\mu\to e\gamma$ and $\mu\to e\overline{e}e$ should be taken into
account.

%%%
\subsection{Dark matter}
\subsubsection{Relic density}
\begin{figure}[t]
\begin{center}
\includegraphics[scale=1]{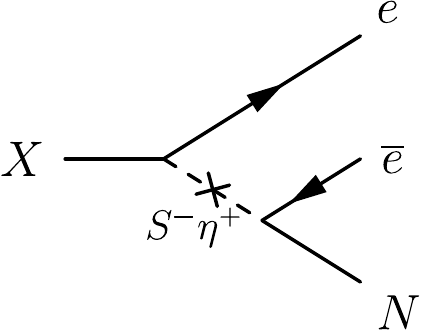}
\caption{Decay process of the metastable particle $X$ via the dimension
 $5$ operator.}
\label{fig:decay}
\end{center}
\end{figure}

We assume that the Majorana fermion $X$ is heavier than $N$ the same as
Model I. 
The $X$ decay process $X\to N e\overline{e}$ is caused by the mixing 
between the charged scalars as depicted in Fig.~\ref{fig:decay}. 
Since the mixing is very small, the fermion
$X$ can have a long lifetime like the previous model. 
However a different point from Model I is that the decaying fermion $X$ is
not a FIMP but a normal WIMP which is thermally produced via the
renormalizable interaction term $y_\eta$. 
First, the DM particle $N$ is produced by the usual freeze-out scenario, then
DM is regenerated by the decay of the metastable fermion $X$ after the freeze-out. 
Consequently a similar situation with Model I can be realized.

The computation of the DM relic density is discussed below. The coupled
Boltzmann equation for $X$ and $N$ is given by~\cite{Fairbairn:2008fb}
%\begin{eqnarray}
%\frac{dn_N}{dt}+3Hn_N&=&-\langle\sigma_N{v}\rangle\left(n_N^2-{n_N^\mathrm{eq}}^2\right)-\Gamma_N n_N,\nn\\
%\frac{dn_X}{dt}+3Hn_X&=&-\langle\sigma_X{v}\rangle\left(n_X^2-{n_X^\mathrm{eq}}^2\right)+N_\mathrm{dec}\Gamma_N n_N,
%\end{eqnarray}
%or
\begin{eqnarray}
\frac{dY_X}{dx}&=&-\frac{s\langle\sigma_{XX}{v_{\rm rel}}\rangle}{xH}
\left[Y_X^2-Y_X^\mathrm{eq2}\left(\frac{m_X}{m_N}x\right)\right]
-\frac{\Gamma_X Y_X}{xH},\nn\\
\frac{dY_N}{dx}&=&-\frac{s\langle\sigma_\mathrm{eff}{v_{\rm rel}}\rangle}{xH}
\Bigl(Y_N^2-Y_N^\mathrm{eq2}\Bigr)
+\frac{\Gamma_X Y_X}{xH},
\label{eq:blt-verII}
\end{eqnarray}
where all the definitions and their values are same with those of the
first model. 
The differential decay width of the decaying particle $X$
for the process $X(p)\to e(k_1)\overline{e}(k_2)N(k_3)$ is calculated as 
\begin{equation}
\frac{d\Gamma_X}{dx_Ed\Omega}(X\to e\overline{e}N)=
m_X\frac{\sqrt{x_E^2-4\xi^2}}{(4\pi)^4}
\frac{\left(1-x_E+\xi^2\right)\overline{\left|\mathcal{M}\right|^2}}
{\left[2-x_E+\sqrt{x_E^2-4\xi^2}\cos\alpha\right]^2},
\label{eq:dwidth}
\end{equation}
where the dimensionless parameters $\xi$ and $x_E$ are defined by $\xi=m_N/m_X$,
$x_E=2E_N/m_X$ with the energy of DM $E_N$ and $\cos\alpha$ is the angle
between the produced DM and the charged lepton in the final state. 
The squared amplitude averaged over initial state spin
is given by
\begin{equation}
\overline{\left|\mathcal{M}\right|^2}=
\frac{2\left|y_Ny_\eta\right|^2\epsilon_2^2v^2{v'}^2\left(p\cdot
	k_1\right)\left(k_2\cdot k_3\right)}
{\left((p-k_1)^2-m_{H_1}^2\right)^2\left((p-k_1)^2-m_{H_2}^2\right)^2}
+\left(k_1\leftrightarrow k_2\right).
\end{equation}
The total decay width $\Gamma_X$ can be obtained by integrating
Eq.~(\ref{eq:dwidth}) in terms of the 
solid angle $\Omega$ and $x_E$ from $2\xi$ to $1+\xi^2$.

For the annihilation cross sections of $X$ and $N$, there are various
annihilation channels such as $XX,NN\to
\ell\overline{\ell},\nu\overline{\nu},q\overline{q},hh,W^+W^-,ZZ$. 
All the channels except the one into the CP-even Higgs bosons are p-wave
dominant which means the cross section 
is proportional to the relative velocity $v_\mathrm{rel}^2$. 
In general, one should include all the channels to compute DM relic
density by solving the coupled Boltzmann equation in
Eq.~(\ref{eq:blt-verII}). 
However in order to find favored parameter space for an interesting gamma-ray signature of
nonthermal DM and to simplify our discussion, it is good to assume $y_X,y_S\ll
y_N,y_\eta$. In this assumption, the 
annihilation cross sections for $X$ and $N$ are extremely simplified and
become p-wave dominant.\footnote{Although more general
discussion with $y_X,y_N\sim y_S,y_\eta$ can be done, this is
phenomenologically less interesting.} 
The coannihilation processes with the charged scalar $S^\pm$ should
also be taken into account 
since we focus on the degenerate case $m_N\approx m_{S^+}$ for sharp
gamma-ray of internal bremsstrahlung. 
Under the assumption $y_X,y_S\ll y_N,y_\eta$, we can use the same
formulas of the (co)annihilation cross sections for DM $N$ with Model I. 
%%%
For the decaying fermion $X$, the main annihilation process is given
by the Yukawa coupling $y_\eta$ and 
there are two channels into a pair of the charged leptons and neutrinos
since the decaying particle $X$ couples with the left-handed lepton
doublet. Thus the cross 
section $\sigma_{XX}{v_{\rm rel}}$ is given by
\begin{eqnarray}
\sigma_{XX}{v_{\rm rel}}&=&
\sigma_{XX}{v_{\rm rel}}(XX\to e\overline{e})+
\sigma_{XX}{v_{\rm rel}}(XX\to\nu\nu)\nonumber\\
&\approx&
\frac{y_\eta^4}{48\pi m_X^2}
\frac{1+\mu_X^2}{\left(1+\mu_X\right)^4}
v_{\rm rel}^2+
\frac{y_\eta^4}{24\pi m_X^2}
\frac{1+\mu_X'^2}{\left(1+\mu_X'\right)^4}
v_{\rm rel}^2,
\label{eq:svX-verII}
\end{eqnarray}
where $\mu_X=m_{\eta^+}^2/m_X^2$ and $\mu_X'=m_{\eta^0}^2/m_X^2$, and
the mass difference between $\eta_R$ and $\eta_I$ is neglected since it
is naturally expected to be small in order to induce the correct
neutrino mass scale. The factor $2$ difference between the two terms in
Eq.~(\ref{eq:svX-verII}) comes from the Majorana nature of the
neutrinos. 
If we consider the degenerate system such as $\mu_X\approx\mu_X'\approx
1$, the coannihilation processes should be taken into account
again. However we do not consider such a case below. 

Note that an additional parameter is required for this model compared to Model
I as one can see from the Boltzmann equations. 
In Model I, the DM relic density is determined by the effective cross section
$\langle\sigma_\mathrm{eff}{v}_\mathrm{rel}\rangle$ and the decay width $\Gamma_X$,
while the cross section 
for the decaying particle $\langle\sigma_{XX}v_\mathrm{rel}\rangle$ is
also needed in Model II. 
Moreover, one more different point of Model II from Model I is that
unlike the FIMP in Model I, the decaying particle $X$ in Model II may 
be detectable by some experiments through the interaction $y_\eta$.

%%%% %%%
\subsubsection{Numerical result}
\begin{figure}[tcb]
\begin{center}
\includegraphics[scale=0.65]{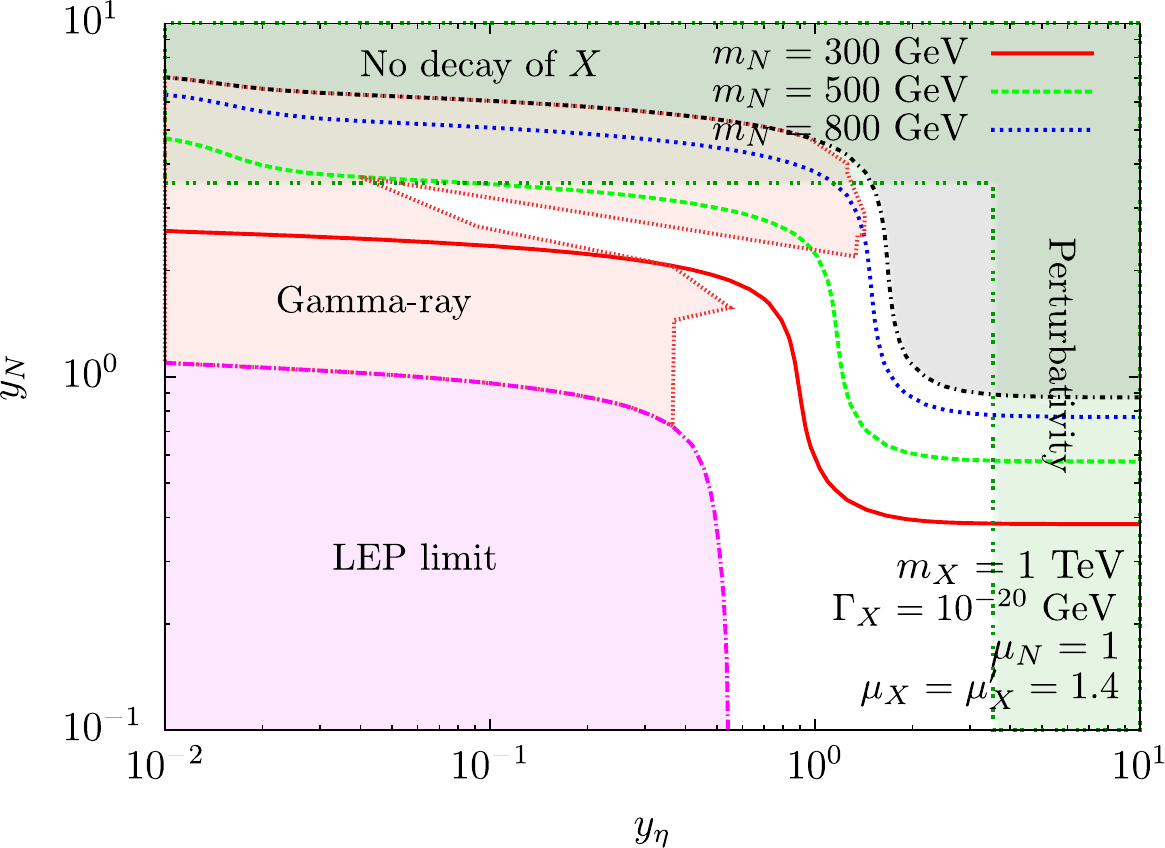}
\includegraphics[scale=0.65]{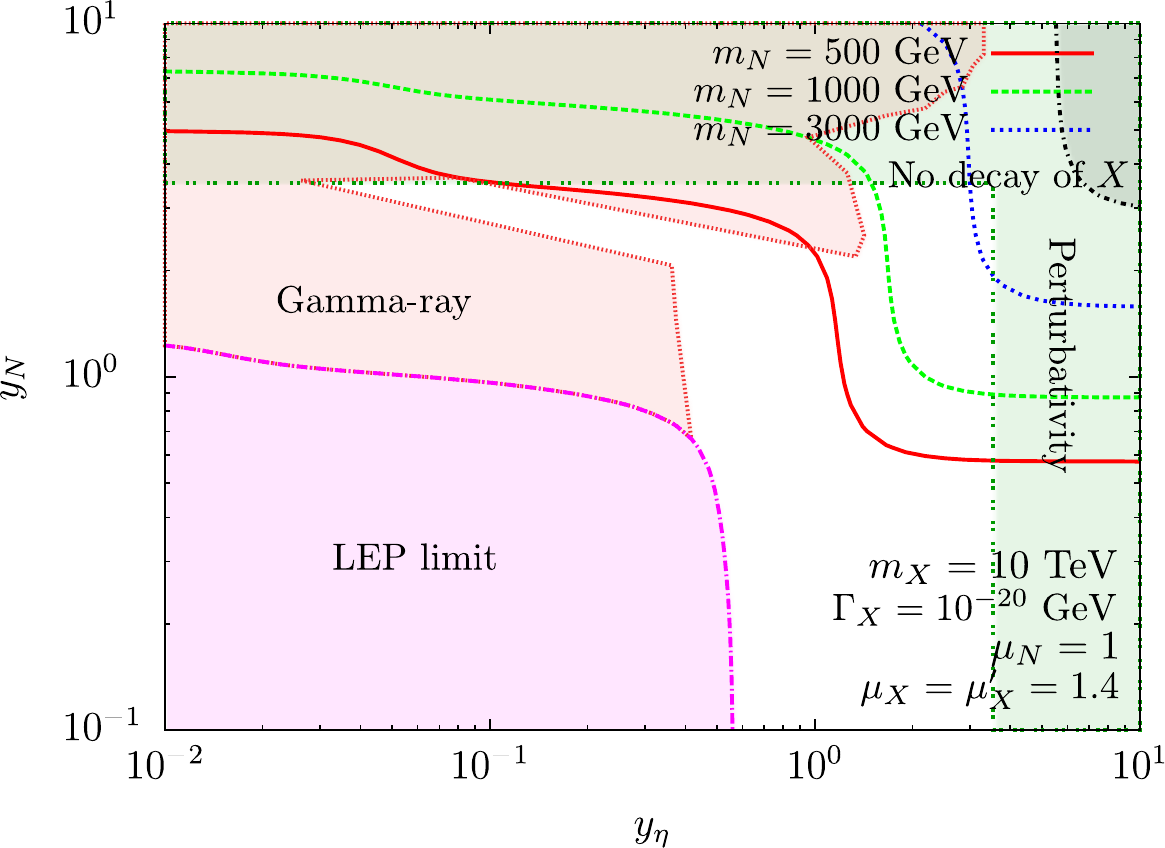}
   \caption{Allowed parameter space in $y_\eta$-$y_N$ plane where the mass of the decaying particle is taken to
 be $m_X=$1 TeV in the left panel and $m_X=10$ TeV in the right one. 
 The same constraints discussed in Model I such as perturbativity, LEP
 and gamma-rays are shown together. 
    Only the white region is allowed by all the current experimental and
 theoretical bounds.} 
   \label{fig:Ver.II}
\end{center}
\end{figure}

The Boltzmann equation Eq.~({\ref {eq:blt-verII}) substituted by Eq
(\ref{eq:eff_sv}) and (\ref{eq:svX-verII}) is numerically solved, and
the result is shown in Fig.~\ref{fig:Ver.II} where the decay width of
$X$ is fixed to be $\Gamma_X=10^{-20}$ GeV, and the mass ratios are
fixed to be $\mu_N=1$ and $\mu_X=\mu_X'=1.4$ to obtain strong sharp gamma-rays. 
The $X$ mass is fixed as $m_X=1$ TeV in the left panel and $m_X=10$ TeV
in the right panel, respectively. 
The basic setup is the same as that in Model I, and only the white region
is allowed by all the current experimental data and theoretical bounds. 

From the figure, one can read off the promising parameter range of $y_\eta$ to see
the interesting gamma-ray signal of nonthermal DM which corresponds to
$y_N\sim\mathcal{O}(1)$ as
\begin{eqnarray}
&&0.5\lesssim y_\eta\lesssim2.0\quad\mbox{for}\quad
m_X=1~\mathrm{TeV},\\
&&0.5\lesssim y_\eta\lesssim3.5\quad\mbox{for}\quad
m_X=10~\mathrm{TeV},
\end{eqnarray}
for $\Gamma_X=10^{-20}$ GeV. 
These ranges are translated to the cross section of the decaying
particle $X$ at the freeze-out times as 
\begin{eqnarray}
&&2.9\times10^{-11}
\lesssim
\frac{\langle\sigma_{XX}v_\mathrm{rel}\rangle}{\mathrm{GeV^{-2}}}
\lesssim 
7.4\times10^{-9}\quad\mbox{for}\quad
m_X=1~\mathrm{TeV},\\
&&2.9\times10^{-13}
\lesssim 
\frac{\langle\sigma_{XX}v_\mathrm{rel}\rangle}{\mathrm{GeV^{-2}}}
\lesssim 
6.9\times10^{-10}\quad\mbox{for}\quad
m_X=10~\mathrm{TeV}.
\end{eqnarray}
One should note that, for a larger cross section
$\sigma_{XX}{v}_\mathrm{rel}$, DM is dominated by thermal production and
is close to the usual WIMP. 
In addition, the cross section for the decaying particle $X$ is also
bounded from above as
$\langle\sigma_{XX}{v}_\mathrm{rel}\rangle\lesssim7.3\times10^{-6}~\mathrm{GeV}^{-2}$
by the perturbativity limit.
Similarly to the case of Model I, deviation from $\mu_N=1$ emerges the
same situation of Model I, but this is beyond our scope.

\section{Conclusions}
\label{sec:3}
From the recent experimental point of view of WIMP searches, the
traditional thermally produced WIMP candidate becomes questionable, and
a different kind of DM is motivated. 
We have proposed two kinds of the models, in which DM relic density is generated
by nonthermal production mechanisms. 
The first model includes FIMPs which can decay into DM. Because of the
existence of FIMPs, DM is 
able to be regenerated after the freeze-out and large couplings of DM
are allowed compared to usual WIMPs. 
In the second model, instead of FIMPs, a thermally produced metastable
particle is able to decay into DM. Then DM relic density can be mainly produced by
the decay of the metastable particle like the first model. 
In addition, the neutrino masses are generated at the one-loop level. 

In these models, we have taken into account some experimental and
theoretical constraints such as the DM relic density, the constraints of
BBN, collider, gamma-rays and perturbativity of couplings. 
We have shown the allowed parameter space of the Yukawa coupling which
can be translated to the DM annihilation cross
section, the decay width of the metastable particle. 
As a feature of nonthermal DM discussed here, a strong indirect
detection signal, especially sharp gamma-rays can be emitted due to
internal bremsstrahlung. This would be a promising channel which is
testable in future gamma-ray experiments such as CTA, GAMMA-400 and DAMPE.

%%%%%%%%%%%%%%%%%%%%%%%%%%%%%%%%%%%
%\vspace{0.5cm}
%\hspace{0.2cm} {\bf Acknowledgments}
\section*{Acknowledgments}
\vspace{0.5cm}
H. O. expresses his sincere gratitude toward all the KIAS members, Korean
cordial persons, foods, culture, weather, and all the other things. 
This work was supported by the Korea Neutrino Research Center which is
established by the National Research Foundation of Korea(NRF) grant
funded by the Korea government(MSIP) (No. 2009-0083526). 
T. T. acknowledges support from P2IO Excellence Laboratory (LABEX).
%%%%%%%%%%%%%%%%%%%%%%%%%%%%%%%%%%%

\end{document}